\documentclass[aps,twocolumn,lengthcheck,preprintnumbers,superscriptaddress]{revtex4-1}%
\usepackage{amsfonts}
\usepackage{amsmath}
\usepackage{amssymb}
\usepackage{graphicx}
\usepackage{color}
\usepackage{txfonts}
\usepackage[unicode]{hyperref}
\hypersetup{
   unicode=true,          
   plainpages=false,
   colorlinks=true,       
   citecolor=blue,        
}

\begin{document}

\author{J. A. Gil Granados}
\affiliation{Departament de F\'isica Qu\`antica i Astrof\'isica, 
Facultat de F\'{\i}sica, Universitat de Barcelona, E--08028
Barcelona, Spain}
\author{A. Mu\~{n}oz Mateo}
\affiliation{Departamento de F\'{i}sica, Universidad de La Laguna, E--38200, La Laguna, Spain}
\author{X. Vi\~nas}
\affiliation{Departament de F\'isica Qu\`antica i Astrof\'isica, 
Facultat de F\'{\i}sica, Universitat de Barcelona, E--08028
Barcelona, Spain}
\affiliation{Institut de Ci\`encies del Cosmos, Universitat de Barcelona, 
ICCUB, 08028-Barcelona, Spain}

\title{Roton instabilities in the superfluid outer core of neutron stars}

\begin{abstract} 
We study the superfluid dynamics of the outer core of neutron stars by
means of a hydrodynamic model made of a neutronic superfluid 
and a protonic superconductor, coupled by both the dynamic entrainment and 
the  Skyrme SLy4 nucleon-nucleon interactions. The resulting nonlinear equations of motion are probed in the search for dynamical instabilities triggered by the relative motion of the superfluids that could be related to observed timing anomalies in pulsars. Through linear analysis, the origin and expected growth of the instabilities is explored for varying nuclear-matter density. Differently from previous findings, the dispersion of linear excitations in our model shows rotonic structures below the pair-breaking energy threshold, which lies at the origin of the dynamical instabilities, and could eventually lead to emergent vorticity along with modulations of the superfluid density. 
\end{abstract}

\maketitle

\section{Introduction}

The study of neutron stars is becoming an interdisciplinary subject 
due to more available data obtained from precise astronomical observations, as well as from more realistic computational simulations.
The neutron-star interior is theoretically described as a layered structure
determined by increasing density and temperature with depth.
Its matter constituents, a mixture of neutron, 
proton, leptons, and eventually another exotic particles, arrange in different manners in concentric shells that are electrically neutral and in $\beta$-equilibrium.
Inside the outermost region, called outer crust, 
matter is organized as a solid lattice of atomic nuclei embedded in a free electron 
gas. This region spans densities from $\rho \sim 10^{4}$ g/cm$^3$, where the 
atomic nuclei are fully ionized, up to $\rho \sim 4\times10^{11}$ g/cm$^3$, where nuclei, which are more and more neutron rich, can not retain the neutrons anymore. At this point, the neutrons start to drip out of the nuclei. In this latter region, called inner crust, 
matter is still arranged as a Coulomb lattice of nuclear clusters, but it is now permeated 
by free neutron and free electron gases. In the bottom 
layers of the inner crust the nuclear clusters may change its shape to minimize 
the Coulomb repulsion from spheres to cylinders, slabs, tubes and spherical or cylindrical bubbles. These
exotic geometries at the crust-core transition are known as "nuclear pasta".
When the density reaches 
$\rho\sim 1.4\times$10$^{14}$ g/cm$^3$, about half the nuclear matter density, 
the solid lattice becomes energetically unfavorable and disappears; as a consequence, at the core, matter transforms into a liquid phase of neutron, protons and leptons.
Deeper in the core, hence at higher densities, strange baryons, and even deconfined quarks may appear. 

At different layers of the neutron star interior,  
the quantum degeneracy of nuclear matter supports the Bose condensation of fermion pairs,  which leads to the emergence of superfluidity. In this way,
neutron pairs are assumed to condense at the inner crust, whereas both neutron and proton pairs can form condensates at the outer core. 
However, the precise equations of state relating the pressure and 
density of nuclear matter under such extreme conditions, including 
pairing correlations, are not completely well determined, and different
proposals compete to provide a plausible picture compatible with the 
observational constrains. 

The observed dynamics of a neutron star is consistent with the assumed superfluid 
interior. The superfluid dynamics can account for the observed low moment of inertia 
\cite{Migdal1960,Baym1969spin,Ravenhall1994,Chamel2012,Watanabe2017} or the 
cooling down of the neutron star 
\cite{Yakovlev2004,Page2004}. It can also play a role in star oscillations and 
collective modes of nuclear matter \cite{Israel2005,Strohmayer2005,Chamel2013}. Additionally,
it allows us to understand the extraordinarily regular rotation of 
pulsars, based on the existence of a superfluid reservoir. 
The sudden speed-ups (or glitches) in this rotation, first detected in the 
Vela and Crab pulsars in 1969 (see \cite{Haskell2015, Fuentes2017} and 
references therein), are one of the most solid arguments in favor of the 
superfluid interior. The theory by P. W. Anderson and N. Itoh 
\cite{Anderson1975} explained the glitches as the result of unpinning quantized 
vortices in the crust. The neutron-superfluid vortices  pinned to the
underlying lattice of nuclei within the inner 
crust, can ``creep'' out of the superfluid transferring angular momentum to the 
non-superfluid part of the star \cite{Alpar1984}. In spite of a general good agreement 
of this theory with observed glitches of different magnitudes, there still 
exist open questions about how different superfluid layers are involved in this phenomenon 
\cite{Chamel2012,Watanabe2017}.  Recent observations (see e.g. 
\cite{Fuentes2017,Ho2017}) seem to favor that the neutron 
superfluid in the inner crust is not enough to provide the observed 
increase in the angular momentum, and suggest that the core superfluids 
have also to be considered. Since many-body analysis of these 
macroscopic, large scale properties are unreachable, most of the studies try 
to model the coupling between different layers of the star interior from a 
hydrodynamical perspective \cite{Chamel2008,Kobyakov2013,Kobyakov2017two,Howitt2016}. 
This approach 
provides the scope for advancing in the understanding of the basic mechanisms 
responsible for dynamical instabilities of the superfluid interior. In turn, such instabilities may shed light on the rotation anomalies of pulsars detected in astronomical observations \cite{Glampedakis2008,Graber2017,Eysden2018,khomenko2019}.

In this context, the present work considers two superfluids coupled by nucleon-nucleon interactions that model the macroscopic structure of the outer core. With typical densities around the  symmetric nuclear matter saturation density $\rho\sim 0.16$ fm$^{-3}$ ($ 2.8\times$10$^{14}$ g/cm$^3$), and temperatures of the order of 10$^8$ K, below plausible estimates for the superfluid transition \cite{Sauls1989}, we assume that the outer core temperature is effectively zero, and therefore that the superfluid density matches the total density of the outer core (see e.g. \cite{Ketterle2008} for the relation between superfluid density and total fermionic density in a fermi gas). 
We have chosen an equation of state (EOS), provided by Douchin and Haensel \cite{Douchin2001}, which is based on the 
Skyrme SLy4 interaction \cite{Chabanat1998}. This EOS is particularly well adapted to the neutron star scenario, and it is compatible with a maximum neutron star mass of about 2 $M_{\odot}$, in agreement with well established observational data \cite{Demorest2010,Antoniadis2013}.
The Galilean invariance of 
the whole system imposes an additional dynamical coupling between the two overlapped condensates  of fermionic pairs due to the entrainment of neutron and protons, whose characteristic quantities are calculated following Ref. \cite{Chamel2006} (see also \cite{Allard2019}).
The resulting model aims at the generalization of previous studies on hydrodynamical instabilities in the coupled superfluids with relative motion \cite{Andersson2004,Kobyakov2017two}. 
To this end,  
the nonlinear Hamilton's equations of the system are linearized, and the dispersion relations and the spectrum of 
excitation modes are analytically derived as a function of the nuclear matter density.
Differently from previous works that have focused on the longest wavelength excitations,  essentially phonons, we show the plausible relevance of rotonic modes developed at shorter wavelengths in the dispersion relations.
 They emerge by virtue of the relative superfluid motion, and are associated with instabilities that can give rise to the decay of the dynamical equilibrium. Such decay involves density modulations, and could eventually lead to emergent quantized vortices. To neatly show the origin of these unstable excitations, we are considering neither the  mutual friction between superfluids, due to the superfluid vorticity inherent in the rotating neutron stars \cite{Alpar1984,Mendell1991},  nor the presence of 
viscosity caused by the underlying normal fluid \cite{Andersson2019}, nor the star magnetic-field, which could have a significant effect on the damping and suppression of linear instabilities \cite{Eysden2018}. 

\section{Model of the superfluid outer core} 

Neutron (denoted hereafter by the subindex $n$) and proton (denoted by subindex $p$) superfluids will 
be described by two real fields, the superfluid densities $\rho_j$ and phases $\theta_j$, with $j=n,p$, where the latter fields provide the irrotational superfluid velocities 
through the relationship $\mathbf{v}_{j}=\hbar\nabla\theta_{j}/m$, with 
$m$ being the nucleon mass. Both fields could also be thought of as real components of effective, bosonic complex order parameters  $\Psi_{j}=\sqrt{\rho_{j}/2}\exp{(i2\theta_{j})}$.
As commonly done in the literature \cite{Mendell1991,Kobyakov2017disp}, we have assumed a zero temperature model where the total density is superfluid, so that $\rho_j$ and $\mu_{j}$ will stand for the fermionic densities and 
chemical potentials of neutrons ($j=n$) and protons ($j=p$), respectively.

The energy density $\mathcal{H}(\rho_n,\rho_p,\nabla\rho_n,\nabla\rho_p,\nabla\theta_n,\nabla\theta_p)$, which 
includes nuclear and Coulomb interactions along with characteristic terms of the superfluid motion, can be recast as
\begin{align}
\mathcal{H}=\mathcal{H}^{\rho}(\rho_n,\rho_p)+\mathcal{H}^{\rho'}(\nabla\rho_n,
\nabla\rho_p)+\mathcal{H}^{\theta'}(\nabla\theta_n,
\nabla\theta_p)+\mathcal{H}^{e}.
\label{eq:Edensity}
\end{align} 
Here,
$\mathcal{H}^{\rho}$ is an homogeneous-density energy term (depending only on the densities) 
provided by the underlaying effective nuclear interaction.
The macroscopic energy density $\mathcal{H}^{\theta'}$ is a dynamical, Galilean-invariant
term that depends on the phase gradients and densities.
The term $\mathcal{H}^{\rho'}$ accounts for density inhomogeneities, and it is a functional of both densities and density gradients.
It contains, on the one-hand, the contributions due to the inhomogeneities of the effective order parameter, and, on the other hand, the contributions of the gradient terms in the nuclear energy-density functional that simulate the finite-range of the nucleon-nucleon interaction.  
Finally, the term $\mathcal{H}^{e}=e\rho_p\Phi/2$ accounts for  the
local charge imbalance through the Coulomb potential $\Phi$, which follows the Poisson equation 
$\nabla^2\Phi=4\pi e(\rho_e-\rho_p)$. The electron density $\rho_e$ is assumed to adjust instantaneously 
into a local density, homogeneous configuration of the ultrarelativistic Fermi gas as $\rho_e=(\mu_e/\hbar c)^3/3\pi^2$. 
Additionally, local $\beta$ equilibrium is assumed, so that the chemical potentials fulfills
the condition $\mu_n=\mu_p+\mu_e$.

The particular functional forms of $\mathcal{H}^{\rho}$, $\mathcal{H}^{\rho'}$, and $\mathcal{H}^{\theta'}$ are 
based on the choice of the phenomenological model for the inter-nucleon interactions (see details below). 
In a generic form, $\mathcal{H}^{\rho'}$ and $\mathcal{H}^{\theta'}$ are given by
\begin{align}
\mathcal{H}^{\rho'}=\dfrac{\hbar^2}{2m}\sum_{ij}\vartheta_{ij}\nabla\rho_{i}
\nabla\rho_{j},
\label{eq:Hgradients1}
\\
\mathcal{H}^{\theta'}=\dfrac{\hbar^2}{2m}\sum_{ij}\varrho_{ij}\nabla\theta_{i}
\nabla\theta_{j},
\label{eq:Hgradients2}
\end{align}
In these equations $\vartheta_{ij}$ and $\varrho_{ij}$, with $i,j=n,\,p$, are density-dependent 
symmetric matrices. We will write the former matrix as $\vartheta_{ij}=\vartheta_{ij}^0+\delta_{ij}/18\rho_j$, where
the constant terms $\vartheta_{ij}^0$ (reported below) are the coefficients of the gradients of the Skyrme energy density functional (see for example Ref. \cite{Chabanat1998}), and the diagonal elements include contributions from the inhomogeneities of the order parameter for each kind of nucleons, 
which can be written in the usual form as $(\nabla \rho)^2/\rho$. Notice that this latter contributions, due to the 
relation between the bosonic and fermionic densities, can also
be understood as the inhomogeneous part of the nuclear kinetic energy density expressed through a diagonal 
Weizs\"acker term \cite{Weizsacker1935}. In Eq.(\ref{eq:Hgradients2}), which corresponds to the dynamical energy density,
the response functions $\varrho_{ij}$ fulfill $\varrho_{nn}+\varrho_{np}=\rho_{n}$, $\varrho_{pp}+\varrho_{pn}=\rho_{p}$, 
and $\varrho_{np}^2<\varrho_{nn}\,\varrho_{pp}$, as independent functions of the velocities \cite{Andreev1976, Chamel2006}. 
Notice that the inhomogeneous energy density (\ref{eq:Hgradients1}) does not contribute to an unperturbed homogeneous
density. but even then it plays an important role when a position-dependent perturbation acts on the system. This is, for instance, the 
case where the crust-core transition density is evaluated using the dynamical method \cite{Gonzalez2019}.

In the absence of a normal fluid (hence without dissipation), the hydrodynamic 
equations of motion can be written following a Hamiltonian approach, based on the energy density Eq. (\ref{eq:Edensity}), 
with canonically conjugate variables $\{\rho_j,\theta_j\}$. The variation of the Hamiltonian $H=\int d\mathbf{r}\,\mathcal{H}$ 
with respect to the densities gives
\begin{subequations}
\begin{align}
 \frac{\partial \rho_{n}}{\partial t}+\nabla\cdot \mathbf{j}_{n}=0 ,\\
 \frac{\partial \rho_{p}}{\partial t}+\nabla\cdot \mathbf{j}_{p}=0,
\end{align}
 \label{eq:continuity}
\end{subequations}
which corresponds to the conservation of particles in each superfluid component. 
Following the seminal paper by Andreev and Bashkin \cite{Andreev1976}, the 
current densities are approximated as linear functions of the 
two superfluid velocities $\mathbf{v}_{n}$ and $\mathbf{v}_{p}$ as:
\begin{subequations}
\begin{align}
\mathbf{j}_{n}= \varrho_{nn}\mathbf{v}_{n}+ \varrho_{np}\mathbf{v}_{p},\\
\mathbf{j}_{p}= \varrho_{pp}\mathbf{v}_{p}+ \varrho_{np}\mathbf{v}_{n}.
\end{align}
\label{eq:current}
\end{subequations}
It is worth noticing that 
$\mathbf{j}_{i}/\rho_i=\mathbf{v}_{i}+(\mathbf{v}_{j}-\mathbf{v}_{i}
)\varrho_{np}/\rho_i$ only matches the corresponding velocity $\mathbf{v}_{i}$ 
if there is no relative motion between the two superfluids.

Additional equations for the potential flow of each superfluid \cite{Mendell1991} follow from the variation of $H$ with respect to the phases:
\begin{subequations}
	\begin{align}
 \frac{\partial \mathbf{v}_n}{\partial t}+
 \nabla\left( \frac{\mu_n }{m} + \mathcal{Q}_n + \frac{|\mathbf{v}_n|^2}{2}-\frac{\partial\varrho_{np}}{\partial\rho_{n}}
\frac{|\mathbf{v}_{pn}|^2}{2} \right) =0,\\
  \frac{\partial \mathbf{v}_p}{\partial t}+
 \nabla\left( \frac{\mu_p}{m} + \frac{e}{m}\Phi+ \mathcal{Q}_p + \frac{|\mathbf{v}_p|^2}{2}-\frac{\partial\varrho_{np}}
{\partial\rho_{p}}\frac{|\mathbf{v}_{pn}|^2}{2} \right) =0,
 \end{align}
  \label{eq:momentum}
 \end{subequations}
where $\mathcal{Q}_j$ are given by
\begin{align*} 
\mathcal{Q}_n=\frac{-\hbar^2}{m^2}\left[\frac{\nabla^2 \sqrt{\rho_n}}{9\sqrt{\rho_n}}  
+\vartheta_{nn}^0 \nabla^2\rho_n + \vartheta_{np}^0 \nabla^2\rho_p \right],
\end{align*}
and equivalently for $\mathcal{Q}_p$ by exchanging the subindexes $n$ and $p$.  
The bulk chemical potentials $\mu_i$ are defined as usual from
\begin{align}
\mu_{n}(\rho_{n},\rho_{p})&=\dfrac{\partial \mathcal{H}^{\rho}}{\partial 
	\rho_{n}},
&  \mu_{p}(\rho_{n},\rho_{p})&=\dfrac{\partial \mathcal{H}^{\rho}}{\partial 
	\rho_{p} }.
\label{eq:mu}
\end{align} 

Equations (\ref{eq:continuity}) and (\ref{eq:momentum}) are the coupled equations of motion for the 
superfluid velocities $\{\mathbf{v}_n, \mathbf{v}_p\}$ and densities $\{\rho_n, \rho_p\}$. 
The system of equations is determined once the nuclear interaction energy density  is known.

 \subsection{Skyrme interaction}
  
In this work we have chosen the phenomenological Skyrme interaction as the underlying nuclear interaction
that describes the non-superfluid EOS \cite{Vautherin1972,Bender2003}.
For this type of effective forces the energy density reads
  \begin{align}
\mathcal{H}^{\mathrm{Skyrme}}=\mathcal{H}^{\rho}+\mathcal{H}^{\nabla},
\label{eq:Hskyrme} \\ 
\mathcal{H}^{\rho}=\mathcal{T}+\mathcal{H}_{0}+\mathcal{H}_{3}+\mathcal{H}_{eff},
\label{eq:Hrho}
\end{align}
where $\mathcal{T}$ is a kinetic term,
$\mathcal{H}_{0}$ is a zero-range two-body term,
$\mathcal{H}_{3}$ is a three-body term, which is recast as a density-dependent, zero-range two-body contribution,
$\mathcal{H}_{eff}$ is an effective-mass term, and $\mathcal{H}^{\nabla}$, which depends on the gradients of the
neutron and proton densities, simulates 
finite-range effects of the interaction and vanishes for uniform density distributions.
Since at the high nuclear density of the outer core the typical Fermi levels are at least one order of magnitude 
greater than the expected superfluid gaps (of the order of one MeV, see below), the pairing contribution to the total energy of the system will be neglected. 

In particular, the first term in Eq. (\ref{eq:Hskyrme}) corresponds to the bulk energy density $\mathcal{H}^{\rho}(\rho_n,\rho_p)$ of Eq. (\ref{eq:Edensity}). 
It gathers the kinetic energy 
densities $\tau_{i}={3}\,(3\pi^{2})^{2/3}\rho_{i}^{5/3}/5$ of both
(nucleonic) Fermi gases in the term $\mathcal{T}$, whereas the next three terms of Eq. (\ref{eq:Hrho}) account for the bulk part of the potential energy density.
The last term in Eq. (\ref{eq:Hrho}), $\mathcal{H}^{\nabla}$, apart from the factor $2m/\hbar^2$, provides the constant $\vartheta_{ij}^0$ coefficients that enter the non-homogeneous energy density
$\mathcal{H}^{\rho'}(\nabla\rho_n,\nabla\rho_p)$ in Eq. (\ref{eq:Edensity}).

The interaction terms are explicitly given by
\begin{align}
\mathcal{T}=\dfrac{\hbar^{2}}{2m}(\tau_{n}+\tau_{p}),
\end{align}
\begin{align}
\mathcal{H}_{0}=\dfrac{t_{0}}{4}\left[(2+x_{0})\rho^{2}-(2x_{0}+1)(\rho_{p}^{2}
+\rho_{n}^{2})\right],
\end{align}
\begin{align}
\mathcal{H}_{3}=\dfrac{t_{3}\,\rho^{{1}/{6}}}{24}\left[(2+x_{3})\rho^{2}-(2x_{3}
+1)(\rho_{p}^{2}+\rho_{n}^{2})\right],
\end{align}
\begin{align}
\mathcal{H}_{eff}&=\left[t_{1}(2+x_{1})+t_{2}(2+x_{2})\right]
\dfrac{\rho(\tau_{n}+\tau_{p})}{8} \nonumber\\&
+\left[t_{2}(2x_{2}+1)-t_{1}(2x_{1}+1)\right]\dfrac{\tau_{p}\rho_{p}+\tau_{n}
\rho_{n}}{8},
\end{align}
\begin{align}
\mathcal{H}^{\nabla}&= \frac{3}{32} [t_{1}(1-x_{1})-t_{2}(1+x_{2})] 
[(\nabla \rho_n)^2 + (\nabla \rho_p)^2]\nonumber\\&
+ \frac{1}{16} [3t_{1}(2+x_{1})-t_{2}(2+x_{2})]\nabla \rho_n \cdot \nabla \rho_p,
\end{align}
with parameters \cite{Chabanat1998,Douchin2001}:
$t_{0}=-$2488.91 MeV fm$^{3}$, $t_{1}=$486.82 MeV fm$^{5}$, $t_{2}=-$546.39 
MeV fm$^{5}$, $t_{3}=$13777.0 MeV fm$^{4}$, $x_{0}=$0.834, $x_{1}=-$0.344, 
$x_{2}=-$1.0, and $x_{3}=$1.354.

\begin{figure}[tb]
	\includegraphics[width=\linewidth]{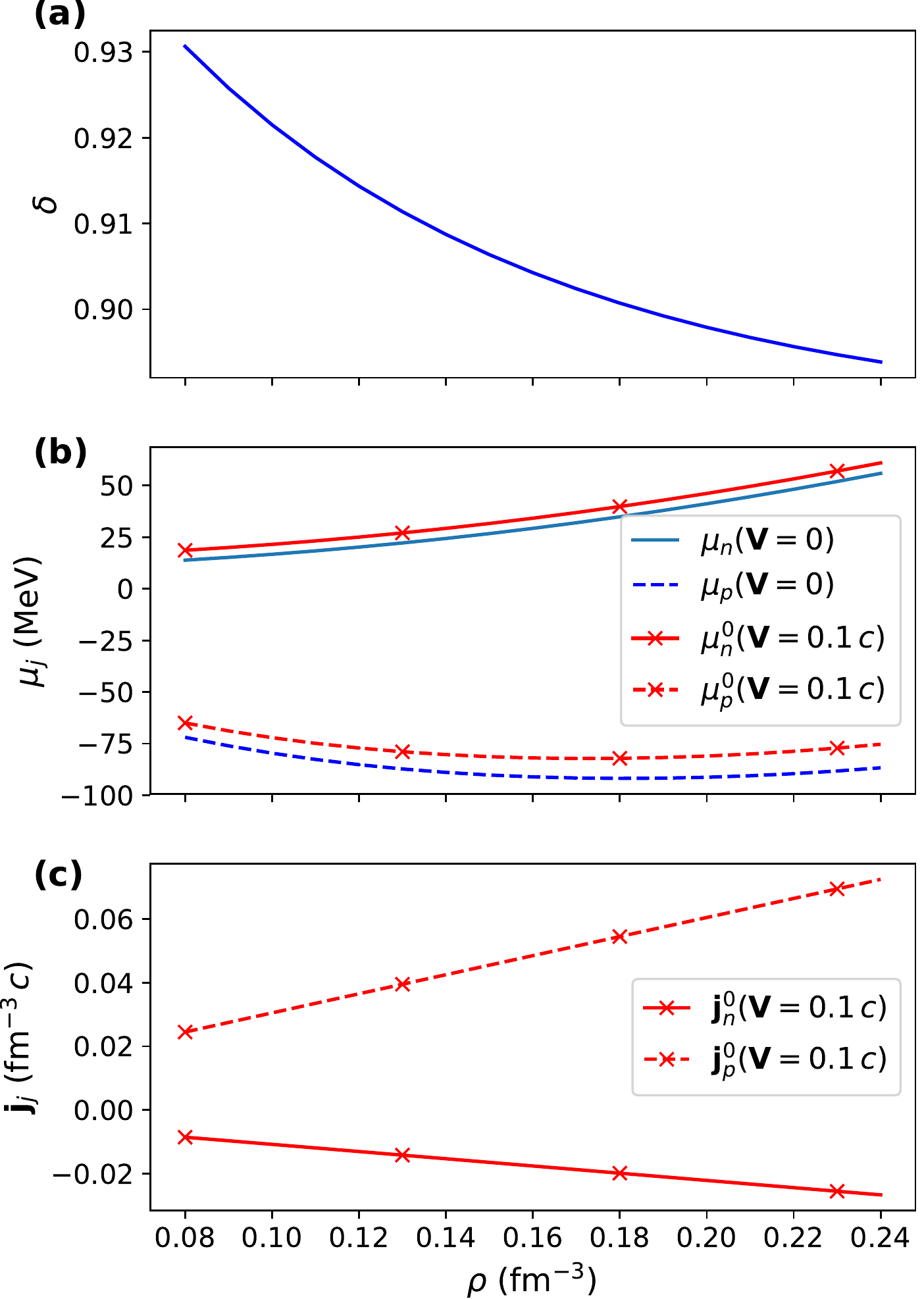}
	\caption{Asymmetric density (a), bulk chemical potential (b), and particle current density (c) as a
		 function of the total density  of nuclear matter in $\beta$ and electric-charge
		  equilibrium, as given by the effective nuclear interaction SLy4. The chemical
	  potentials are represented for static and dynamical equilibrium with relative
  velocity $2\mathbf{V}=0.2\,c$ between proton and neutron superfluids.}
	\label{fig:beta}
\end{figure}

By means of Eq. (\ref{eq:Hskyrme}), the $\beta$ equilibrium  in the bulk can be locally determined 
in terms of the asymmetric density $\deltaup=({\rho_{n}-\rho_{p}})/{\rho}$ by the condition $\mu_n=\mu_p+\mu_e$, as:
\begin{align}
\left(\dfrac{\hbar^{2}}{2m}+ \frac{C_1}{8}\rho\right)\left(3\pi^{2} \dfrac{\rho}{2}\right)^{2/3} 
\left[(1+\deltaup)^{2/3}-(1-\deltaup)^{2/3}\right] \nonumber\\
-\dfrac{t_{0}}{2}\rho\,\deltaup\,(1+2x_{0}) 
-\dfrac{t_{3}}{12}(2x_{3}+1)\deltaup\,\rho^{7/6} \nonumber\\
+\dfrac{3\pi^{2}}{5}C_2\left(\dfrac{\rho}{2}\right)^{5/3}
\left[(1+\deltaup)^{5/3}-(1-\deltaup)^{5/3}\right] \nonumber\\
-\hbar c\left(3\pi^2\rho_{e}\right)^{1/3}=0 ,
\label{eq:beta}
\end{align}
where the coefficients $C_1=t_{1}(2+x_{1})+t_{2}(2+x_{2})$ and $C_2=t_{2}(2x_{2}+1))-t_{1}(2x_{1}+1)$, which enter the Skyrme energy density
through $\mathcal{H}_{eff}$, are related to the neutron and 
proton effective masses by
\begin{align}
\left(\frac{m}{m^*}\right)_n = 1 + \frac{2m}{\hbar^2}[C_1\rho + C_2\rho_n], \\
\left(\frac{m}{m^*}\right)_p = 1 + \frac{2m}{\hbar^2}[C_1\rho + C_2\rho_p]. \nonumber 
\end{align}

The numerical solution to Eq. (\ref{eq:beta}) is represented in the top panel of Fig. 
\ref{fig:beta} for static conditions and assuming local electric charge balance $\rho_e=\rho_p$. 
As can be seen,  in the case of the SLy4 interaction the asymmetry decreases for increasing nuclear density. 
The corresponding chemical potentials are depicted in the middle panel of Fig. \ref{fig:beta}.
 For large asymmetries, well above its neutron drip value, the neutron chemical potential is positive, owing to
the repulsive character of the neutron-neutron interactions, whereas the proton chemical potential is strongly negative due to
the attractive neutron-proton interaction (see \cite{Centelles1998} for more details).

The response function $\varrho_{np}$, which  determines the entrainment, is also obtained consistently with the
help of the same SLy4 effective force used to describe the nucleon-nucleon interaction. Following  Ref. \cite{Chamel2006} one gets
\begin{equation}
 \varrho_{np}=\alpha\,\rho_{n}\rho_{p},
\end{equation}
 where the constant $\alpha$ depends on the parameters of the Skyrme force as
\begin{equation}
 \alpha=-\dfrac{m}{2}\left[ t_{1}\left( 1+\dfrac{x_{1}}{2}\right) 
 +t_{2}\left( 1+\dfrac{x_{2}}{2}\right) \right],
\end{equation}
 which in the case of the SLy4 interaction takes the value $\alpha \approx -1.566$ fm$^{3}$.

 \section{Stationary states}
 
The superfluid Eqs. (\ref{eq:continuity}) and (\ref{eq:momentum}) admit constant density $\{\rho_n^0, \rho_p^0\}$ 
stationary solutions with constant relative velocity $\mathbf{v}_{pn}^0=\mathbf{v}_{p}^0-\mathbf{v}_{n}^0$.
For the sake of simplification, we choose superfluid velocities with equal modulus and opposite direction 
$\mathbf{v}_{p}^0=-\mathbf{v}_{n}^0= \mathbf{V}$,  which implies
$\mathbf{v}_{pn}^0=2\mathbf{V}$.
For these states, the total chemical potentials  at equilibrium $\{\mu_n^0, \mu_p^0\}$ are 
\begin{equation}
\begin{aligned}
\mu_n^0 = \mu_n(\rho_n^0, \rho_p^0)+  \frac{m|\mathbf{V}|^2}{2}\left(1-4\alpha\rho_p^0\right) ,\\
\mu_p^0 = \mu_p(\rho_n^0, \rho_p^0)+  \frac{m|\mathbf{V}|^2}{2}\left(1-4\alpha\rho_n^0\right),
\label{eq:steady}
\end{aligned}
\end{equation}
and the corresponding particle current densities
\begin{equation}
\begin{aligned}
\mathbf{j}_n^0 = -(\rho_n^0-2\alpha \rho_p^0)\mathbf{V}  , \\
\mathbf{j}_p^0 = (\rho_p^0-2\alpha \rho_n^0)\mathbf{V},
\label{eq:Jsteady}
\end{aligned}
\end{equation}
so that the total current density is $\mathbf{j}^0=-(1+2\alpha)\, \deltaup^0\,\rho^0\,\mathbf{V}$.
To give a more concrete idea of the influence of the relative superfluid velocity,  we have represented the equilibrium quantities for $\mathbf{V}=0.1\,c$ in the middle and lower panels of Figure \ref{fig:beta}. 
The increase in the chemical potential with respect to the static case ($\mathbf{V}=0$) is around 10$\%$ for protons and 17$\%$ for neutrons at saturation density. The chemical potential of electrons follows from the local charge and $\beta$ equilibrium.

\subsection{Linear excitations}
The dynamical stability of the stationary states can be studied by analyzing the spectrum of linear excitations. To do so, we introduce a perturbation on the equilibrium state, as $\{\rho_j,\,\mathbf{v}_j\}\rightarrow \{\rho_j^0 +\delta \rho_j,\,\mathbf{v}_j^0 +\delta \mathbf{v}_j\} $, substitute the perturbed quantities in the equations of motion, Eqs. (\ref{eq:continuity}) and (\ref{eq:momentum}), and keep terms up to linear order in the perturbations. In matrix form, the resulting linearized equations read 
\begin{align}
\frac{\partial U}{\partial t}  = -\nabla\left(\mathcal{B} U\right)
\label{eq:linear_sys}
\end{align}
 where $U$ is the vector of perturbations $[\delta \rho_n,\,\delta \rho_p,\,\delta \mathbf{v}_n,\,\delta \mathbf{v}_p]^T$, and the linear operator $\mathcal{B}$ is given in terms of the equilibrium quantities  by
\begin{align}
\mathcal{B}=
\left(\begin{array}{cccc} 
\frac{\mathbf{j}_{n}}{\rho_n} &
\frac{\varrho_{np}}{\rho_p}\mathbf{v}_{pn} & \varrho_{nn} & \varrho_{np} \\
-\frac{\varrho_{np}}{\rho_n}\mathbf{v}_{pn} & \frac{\mathbf{j}_{p}}{\rho_p}   &
 \varrho_{np} &  \varrho_{pp}\\
 \frac{g_{nn}}{m} & \frac{g_{np}}{m}-\frac{\alpha}{2}\mathbf{v}_{pn}^2 &
  \frac{\mathbf{j}_{n}}{\rho_n}  & -\frac{\varrho_{np}}{\rho_n}\mathbf{v}_{pn} \\
 \frac{g_{pn}}{m}-\frac{\alpha}{2}\mathbf{v}_{pn}^2 &
 \frac{g_{pp}+e\Phi}{m}& \frac{\varrho_{np}}{\rho_p}\mathbf{v}_{pn} & \frac{\mathbf{j}_{p}}{\rho_p}
\end{array}\right),
	\label{eq:linearB}
\end{align}
where the stationary ($^0$) superscripts have been omitted in the matrix elements for the sake of uncluttered expressions. We have introduced the operators
\begin{align}
g_{ij}&=\dfrac{\partial\mu_{i}}{\partial\rho_{j}}-\frac{\hbar^2}{2m}\vartheta_{ij}\nabla^2, 
\label{eq:Eij}
\end{align}
which gather interaction and dispersion contributions. The former contribution
plays the role of an interaction strength, either
intra-component $g_{nn}$ and $g_{pp}$, or inter-component $g_{np}=g_{pn}$,
reflecting the density coupling between neutrons and protons. The response of the electrons in the absence of collisions
is accurately provided by the random phase approximation, which for ultrarelativistic electrons is given by
$4\pi e\delta\rho_e = k^2(\varepsilon_e-1) \delta\Phi$, with dielectric constant  $\varepsilon_e(k)\sim 1+4\pi e^2(3\,\rho_e/\pi)^{2/3}/(\hbar c k^2)$ at zero frequency \cite{Jancovici1962,Kobyakov2017disp}. This electronic response applies for small wavenumbers and energies against the corresponding electronic Fermi values; for instance, at $\rho=0.16$ fm$^{-3}$ and a proton fraction of $\rho_p/\rho=0.05$, this means  $k< 0.61$ fm$^{-1}$ and $\hbar\omega < 120$ MeV. The zero frequency assumption is justified whenever $\omega\ll kc$,  which lies within the error margins of the present model for the typical modes that we find in our calculations (up to $\omega\sim 0.3\,kc$).
According to this prescription, in Eq. (\ref{eq:linearB}) we have made the approximation $e\Phi\approx 4\pi e^2/(\varepsilon_e \,k^2)$. 
For later use in the long wavelength limit $k\rightarrow 0$, we also define the quantity $g_{ee}=\partial \mu_e/\partial \rho_e=(3\rho_e/\pi)^{2/3}/(\hbar c)$, which  
is related to the compressibility $\kappa$ of the ultrarelativistic Fermi gas by the relation $\kappa^{-1}=g_{ee}\,\rho_{e}$. Overall, we neglect the electronic damping associated to the imaginary part of the dielectric constant ($\mathrm{Im}[\varepsilon_e]\propto \omega\,k^{-3}$) \cite{Jancovici1962}, since it is expected to have significant effects at much smaller wavenumbers (e.g. $k<0.1$ fm$^{-1}$ at $\rho=0.16$ fm$^{-3}$) than the relevant wavenumbers for instabilities considered here. 

\begin{figure}[tb]
	\includegraphics[width=\linewidth]{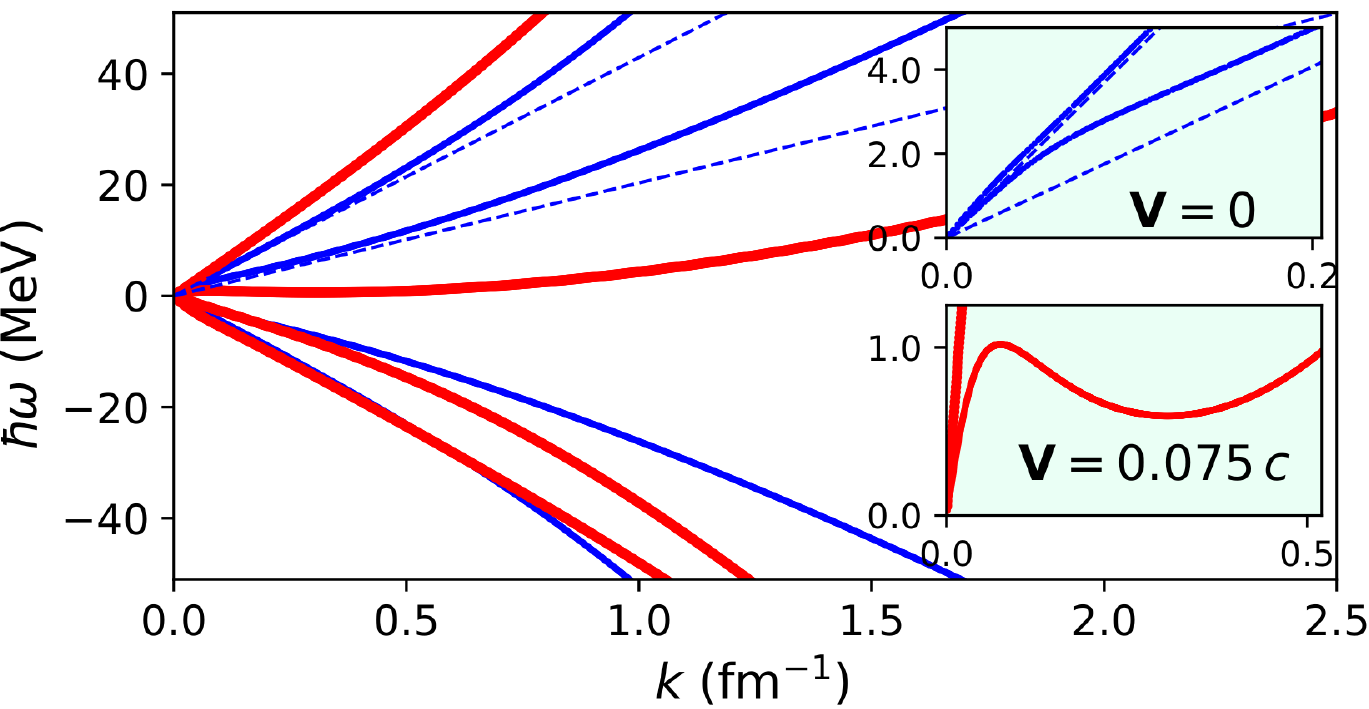}\\
	\vspace{0.2cm}
	\includegraphics[width=0.97\linewidth]{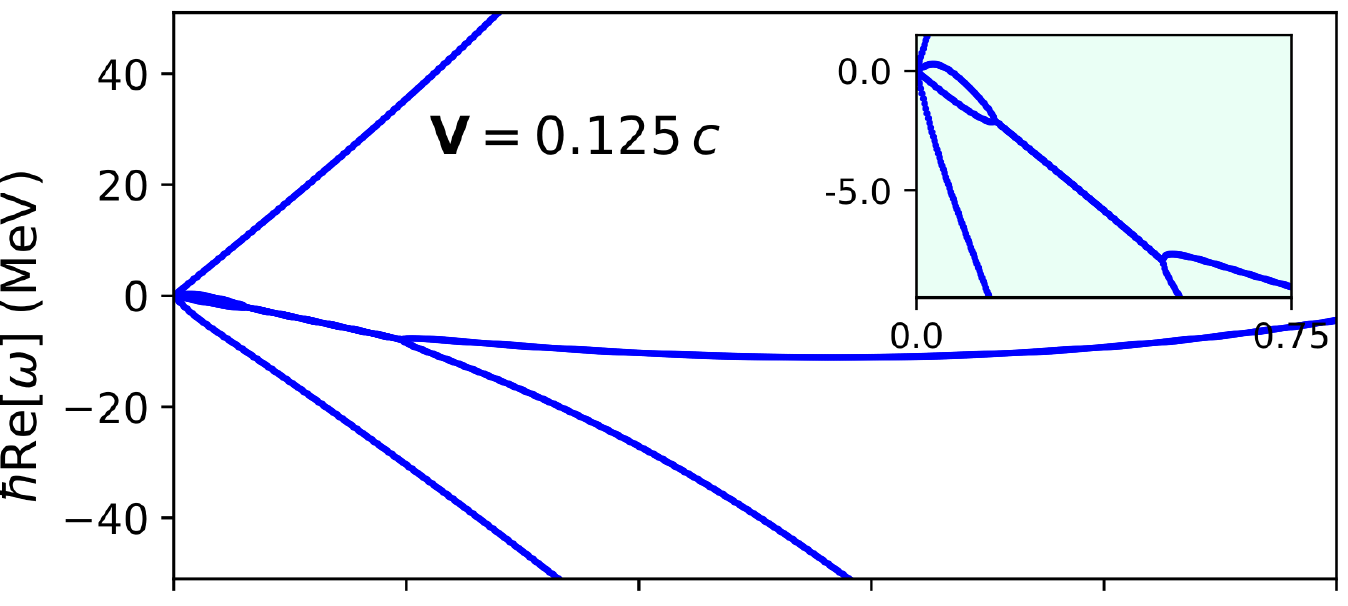}\\
	\includegraphics[width=\linewidth]{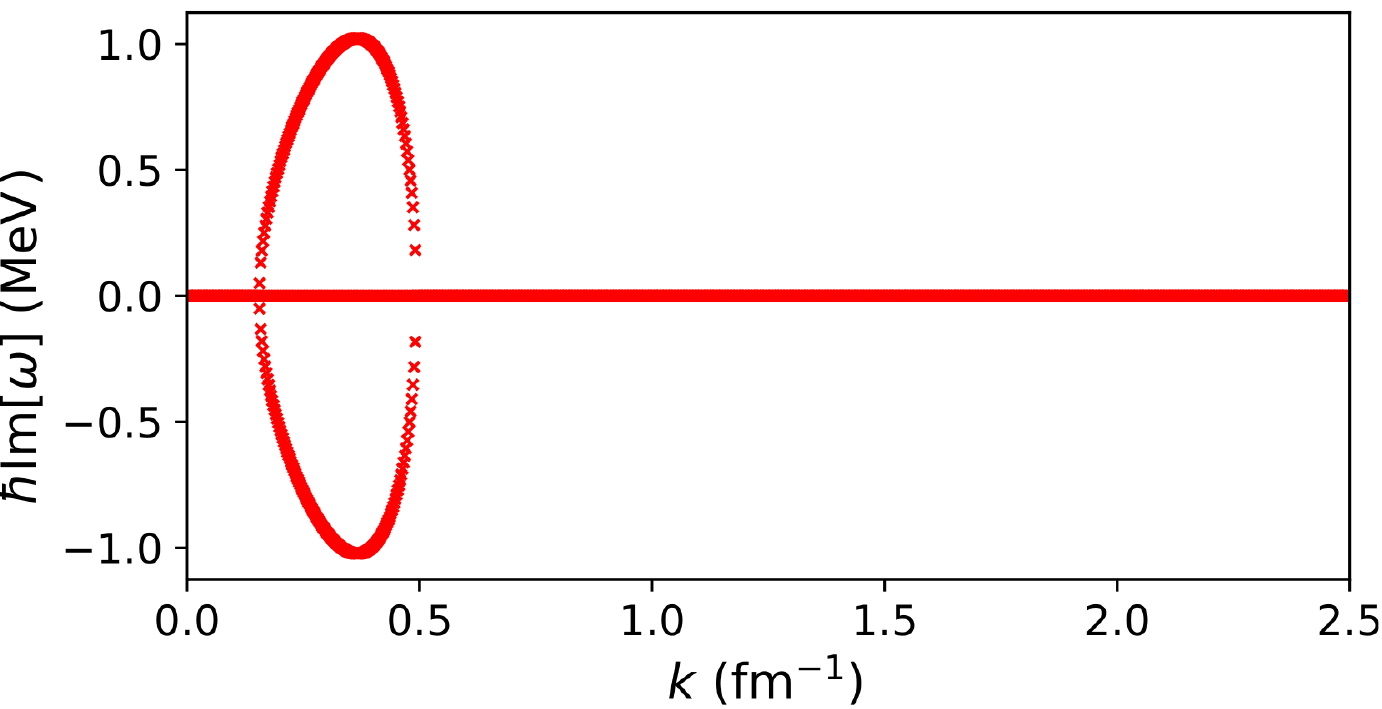}
	\caption{Linear excitation energies in a uniform state with total density $\rho=0.16$ fm$^{-3}$ and several relative superfluid velocities $\mathbf{v}_{pn}=2\mathbf{V}$. The dispersion contains only stable, real frequencies for $\mathbf{V}=0$, and $0.075\,c$ (top panel), whereas it includes complex frequencies (unstable excitations) for $\mathbf{V}=0.125 \,c$ (middle and bottom panels). The dashed lines in the top panel trace the speed of sound of superfluid neutrons (higher slope) and superfluid protons (lower slope) when considered alone (see text). The insets show close-ups of the rotonic regions, below (top-panel inset) and beyond (middle- panel inset) the relative velocity threshold for the onset of instabilities.}
	\label{fig:freq016}
\end{figure}

The translational invariance of the matrix (\ref{eq:linearB}) allows for the expansion of the perturbation in plane waves 
$U(\mathbf{r},t)=\sum_\mathbf{k} U_\mathbf{k} \exp[i(\omega t-\mathbf{k} \cdot \mathbf{r})]$, and so the dispersion relations 
$\omega(\mathbf{k})$ can be readily obtained as the algebraic solution of the linear system 
(\ref{eq:linear_sys}). Notice that this system consists of two scalar equations, for the density perturbations. and two vector equations, for the velocity perturbations; therefore, for general vector perturbations, the excitation frequencies $\omega$ are the solutions of an eigenvalue problem with a characteristic eighth order polynomial. 
Since we are interested in the dynamical effects of the relative velocity, from now on we will focus on density and velocity perturbations along the direction of the relative velocity, that is, with wavenumber $ k=\mathbf{k}\cdot\mathbf{V}/\mathrm{V}$, and neglect the perturbations in the perpendicular direction. 
In this case, the dispersion relations derived from Eq. (\ref{eq:linear_sys}) are obtained as the roots of a fourth order characteristic polynomial in the variable $\omega/k$. 
The dispersion shows two qualitatively different regions, a phonon part at low $k$ where $\omega\propto k$, and a particle part at high $k$, where $\omega\propto k^2$ (see the top panel of Fig. \ref{fig:freq016}). Although these features are typical of the spectrum of linear excitations in a gas of condensed bosons \cite{Pitaevskii2003}, the underlying Fermi gases pose additional constraints on the excitation of collective modes. It is well known that in the presence of fermionic pairing there exist bosonic excitations at low wavenumber (phonons), the so-called Bogoliubov-Anderson modes \cite{Anderson1958}, which lay (in a system of electrically neutral particles) within the superfluid gap.  
But the Fermi gases present also fermionic (quasi-particle) excitations that lead to fermion-pair breaking, found typically at twice the gap energy. Therefore, the pair-breaking excitations represents an energy threshold beyond which the collective modes  are damped \cite{Martin2014} (see also \cite{Combescot2006} for a discussion in ultracold gases). 
It is then crucial to account for the typical values of the superfluid energy gaps in order to determine the undamped collective modes below the mentioned threshold. 

We estimate the available energy range for collective excitations from the calculated 
nucleonic $^1 S_0$ and $^3 PF_2$ pairing gaps of Ref. \cite{Zhou2004}, for protons and neutrons respectively, which are calculated at BCS level with single-particle energies obtained with the Br\"uckner-Hartree-Fock approach including two-body (Argonne $v_{18}$) and three-body (Urbana UIX) forces. 
In the range of nuclear densities of the outer core, the neutron $^3 PF_2$ pairing increases approximately with a linear slope $0.23$ MeV $\times$ fm$^3$ up to $\sim 0.8$ MeV at $\rho=0.24$  fm$^{-3}$. The proton $^1 S_0$ pairing shows a non-monotonic behavior (approximately as an inverted parabola) that reaches a maximum of $\sim 0.8$ MeV  close to nuclear saturation density (see also \cite{Baldo2007,Guo2019}). 
According to these values, the relevant collective modes of Fig. \ref{fig:freq016}, for $\rho=0.16$ fm$^{-1}$ and neutron energy gap $\sim 0.4$ MeV (see Fig. 2c in Ref. \cite{Zhou2004}), are those below the pair-breaking energy limit $|\hbar\omega|\lesssim 0.8$  MeV. 
Notice that these thresholds apply to static systems ($\mathbf{V}=0$), whereas in moving systems they have to be measured [as the chemical potentials in Eq.( \ref{eq:steady})] with respect to the kinetic-energy shifted Fermi levels; for instance, for $\mathbf{V}=0.125$ c, the  kinetic-energy shift amounts to 4.9 MeV in neutrons and 9.2 MeV in protons. 
Although the particle-like region of the spectrum of collective excitations is effectively suppressed above this threshold by the pair-breaking excitations, the lowest energy modes remain undamped. Within this energy window,  along with phonons, as we show below, roton-like modes can be found at intermediate wavelengths.

\subsubsection{Simplified linear models}

 Before discussing the general dispersion, it is instructive to look first at simple cases of phonon modes. 
For the static equilibrium ($\mathbf{v}_j=0$) in the absence  of entrainment ($\varrho_{np}=0$), the continuity Eqs. (\ref{eq:continuity}) are decoupled, whereas the momentum Eqs. (\ref{eq:momentum}) are coupled by density terms ($g_{np}(k=0)<0$ for $\rho<0.21$ fm $^{-3}$); then,
within the long wavelength limit $k\rightarrow 0$, the dispersion can be approximated by
\begin{align}
{\omega}_{(\varrho_{np}=0)}=\pm k \left[\frac{ c_{n}^2+{c_{p}^*}^2}{2}  
\pm\sqrt{\left(\frac{c_{n}^2-{c_{p}^*}^2}{2}\right)^2 + c_{np}^4}\,\,\right]^\frac{1}{2}
\label{eq:sound0}
\end{align}
where $c_n =\sqrt{g_{nn}\,\rho_{n}/m}$ and $c_p =\sqrt{g_{pp}\,\rho_{p}/m}$ are respectively the speed of sound in the neutron 
and proton superfluids (when considered alone), and $c_{np} =\sqrt{|g_{np}|\sqrt{\rho_{n} \,\rho_{p}}/m}$ is an analogous speed 
term associated with the density coupling. 
The effective speed of sound for protons $c_p^*=\sqrt{c_p^2+g_{ee}\,\rho_{p}/m}$, assuming charge equilibrium, includes an electronic contribution accounted for by means of $g_{ee}$.
For example, at $\rho=0.16$ fm$^{-3}$, one gets $c_n =0.22\, c$, $c_p =0.1\, c$, and $c_p^* =0.23\, c$; the resulting speed of sound of the excitations is 
$c_{(\varrho_{np}=0)}=\omega_{(\varrho_{np}=0)}/k=\pm 0.24\, c$ for the top and bottom branches of the dispersion, and 
$\pm 0.21 \, c$ for the intermediate branches. These values reflect a high influence of the electron gas at very low wavenumber, 
typically at $k<0.1$ fm$^{-1}$; beyond this region, the dispersion branches show a tendency that is better captured  by slopes given by $c_n$, and $c_p$ (represented by dashed lines in the top panel of Fig. \ref{fig:freq016}). 

In the excitation spectrum associated with the asymptotic dispersion (\ref{eq:sound0}), the velocity components follow the density component of the modes as $\delta \mathbf{v_p}/\delta \mathbf{v_n}=(\rho_n/\rho_p)\,\delta\rho_p/\delta\rho_n$, which can show a (proton to neutron) in-phase character $\delta\rho_p/\delta\rho_n\,>0$, corresponding to the lower (in absolute value) energy branches, or an out of phase 
character $\delta\rho_p/\delta\rho_n<0$, corresponding to the higher energy branches in the top panel of Fig. \ref{fig:freq016}. 
Notice that all the stable modes (having pure real frequencies in the dispersion) can be chosen to have only real components, 
so $U_\mathbf{k}\in \mathbb{R}$ (see Fig. \ref{fig:mod016} for the complete spectrum 
without approximation and varying wavenumber). Conversely, the unstable modes can be complex $U_\mathbf{k}\in \mathbb{C}$, although the real 
constraint on the perturbed quantities implies $U_\mathbf{k}=\pm U_\mathbf{-k}^*$.
 
On the other hand, when the density coupling in the energy density functional is neglected ($g_{np}=0$), so that the momentum Eqs. (\ref{eq:momentum}) become decoupled whereas the entrainment keeps the continuity Eqs. (\ref{eq:continuity}) coupled, the static equilibrium leads to the following dispersion of linear excitations in the phononic limit $k\rightarrow 0$:
\begin{align}
{\omega}_{(g_{np}=0)}=\pm k \left[\frac{ c_{nn}^2+{c_{pp}^*}^2}{2}  
\pm\sqrt{\left(\frac{c_{nn}^2-{c_{pp}^*}^2}{2}\right)^2 + c_\alpha^4 }\,\,\right]^\frac{1}{2}
\label{eq:sound1}
\end{align}
where now $c_{nn} =\sqrt{g_{nn}\,\varrho_{nn}/m}$, and $c_{pp}^* =\sqrt{c_{pp}+g_{ee}\,\varrho_{pp}/m}$, with $c_{pp} =\sqrt{g_{pp}\,\varrho_{pp}/m}$, are the speeds of sound modified 
by the entrainment, and $c_\alpha =\sqrt{ \sqrt{g_{nn}\,(g_{pp}+g_{ee})}\,\varrho_{np}/m}$ is a specific speed term associated with 
the entrainment coupling. From examination of Eqs. (\ref{eq:sound0}) and (\ref{eq:sound1}), one can see that both types of coupling, 
either density or entrainment coupling, produce qualitatively similar type of phonon excitations.  Again, the velocity components follow the density components of the modes as per $\delta \mathbf{v_p}/\delta \mathbf{v_n}=(c_{pp}/c_{nn})^2(\rho_n/\rho_p)\,\delta\rho_p/\delta\rho_n$, with the in-phase (respectively, out of phase) modes corresponding to the lower (resp. higher) energy branches.
For concreteness, at $\rho=0.16$ fm$^{-3}$, the speeds of sound are $c_{(g_{np}=0)}=\omega_{(g_{np}=0)}/k=\pm 0.26\, c$ and $\pm 0.22 \, c$ 
for the four branches. At low wavenumber $k<1$, but beyond the zone of influence of the electron gas, the dispersion in the absence of density coupling is well approximated by linear branches of slope $c_{nn} =0.22\, c$, and $c_{pp} =0.11\, c$. Note that the presence of entrainment changes neither quantitatively the phonon velocities, since $c_{nn}\approx c_n$ and $c_{pp}\approx c_p$, nor qualitatively the dispersion branches, which are monotonically growing in energy for increasing wavenumber.

Finally, if both entrainment and density coupling are neglected, but relative superfluid 
velocity $2\mathbf{V}$ is considered, the dispersion at $k\rightarrow 0$ shows
shifted sound speeds
${\omega}_{(\varrho_{np}=0,\,g_{np}=0)}= k \,(\pm c_{n}-\mathbf{V})$ and ${\omega}_{(\varrho_{np}=0,\,g_{np}=0)}= k \,(\pm c_{p}^*+\mathbf{V})$, as corresponds to decoupled superfluids.
For comparison with the previous cases, at $\rho=0.16$ fm$^{-3}$ and $\mathbf{V}=0.1\,c$ one gets $c_{(\varrho_{np}=0,\,g_{np}=0)}=-0.32 \, c$ and $0.12  \, c$ for neutrons, and $0.33 \, c$ and $-0.13  \, c$ for protons.
The effect of the relative velocity on the whole dispersion tends to separate the higher (in absolute value) energy  branches, and to get closer the 
lower energy branches (see top panel of Fig. \ref{fig:freq016}); the latter ones can eventually collide and, as a result, produce instabilities.

\subsubsection{Full linear model}
In the general case, by combining density and entrainment coupling with a relative superfluid velocity, the plane-wave solutions to the 
linear system Eq. (\ref{eq:linear_sys}) gives rise to a generic fourth order polynomial for the excitation frequencies. Its cumbersome 
analytical solution is however easily obtained numerically.
Figure \ref{fig:freq016} shows the linear-excitation frequencies for total nuclear matter density $\rho=0.16$ fm$^{-3}$ and several 
relative velocities. The top panel compares the excitations of the static equilibrium $\mathbf{V}=0$ and the dynamical equilibrium at  
$\mathbf{V}=0.075 \,c$. The symmetric branches (with respect to the zero frequency) at $\mathbf{V}=0$ contrast with the asymmetric 
dispersion in the presence of relative velocity. The dispersion of the static equilibrium is characterized by a phonon region at low wavenumber $k\rightarrow 0$, which can be reasonably approximated by Eq. (\ref{eq:sound0}). 
In the dynamical equilibrium, as can be better seen in the inset, the lower (in absolute value) 
energy branches bend towards each other, and develop a rotonic structure with a minimum phase velocity $\omega/k$ at nonzero momentum, as can be seen for $\mathbf{V}=0.075 \,c$.  
Further increase in the relative velocity of the system, as shown in the lower panels of Fig. \ref{fig:freq016} for $\mathbf{V}=0.125 \,c$, 
leads to a dynamical instability associated to complex excitation energies. The unstable region covers a limited range of intermediate wavenumbers  $k\in  [0.16,\,0.49]$ fm$^{-1}$ (see the bottom panel of Fig. \ref{fig:freq016}) between mode bifurcations. The growth rate of these instabilities is expected to be 
exponential during small times (roughly until the moment when the perturbations amplitudes reach values of the same order as the equilibrium quantities), 
and the typical time for their development  $\tau$ is inversely proportional to the maximum value of the imaginary frequencies $\tau\propto$ Im$[\omega]^{-1}$. 
For the unstable case depicted in 
the middle ant bottom panels of Fig. \ref{fig:freq016}, this linear time estimate is $2\pi\omega_{max}^{-1}= 4.1\times 10^{-21}$ s. Notice that this value (calculated in the linear regime) is not the time it takes for the superfluids to decay from their equilibrium state, which is a subsequent nonlinear process. For comparison, it is worth pointing out that the linear estimate is 2.4 times longer than the period of the proton plasmon  $\tau_p = 1.7\times 10^{-21}$ s (as calculated for decoupled protons).

\begin{figure}[tb]
\flushleft{ \bf{(a)}}

\centering
	\includegraphics[width=0.95\linewidth]{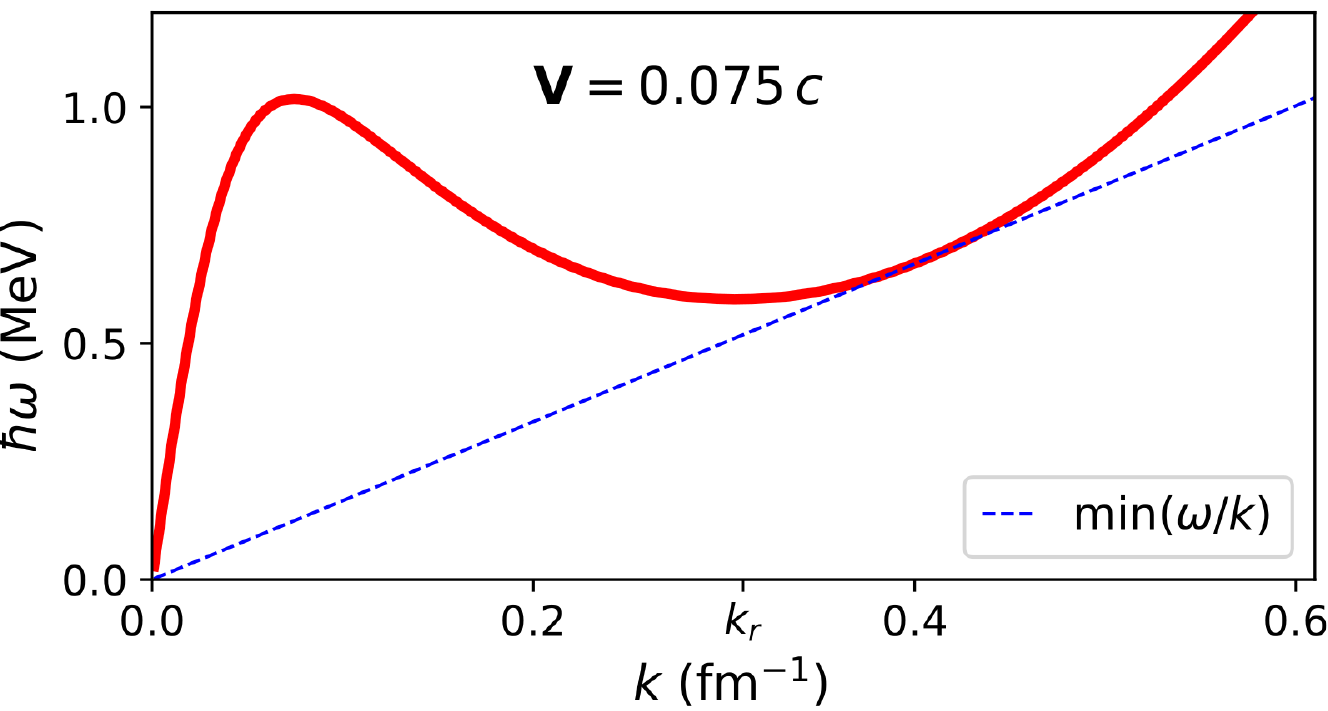}
\flushleft{ \bf{(b)}} 
	\includegraphics[width=\linewidth]{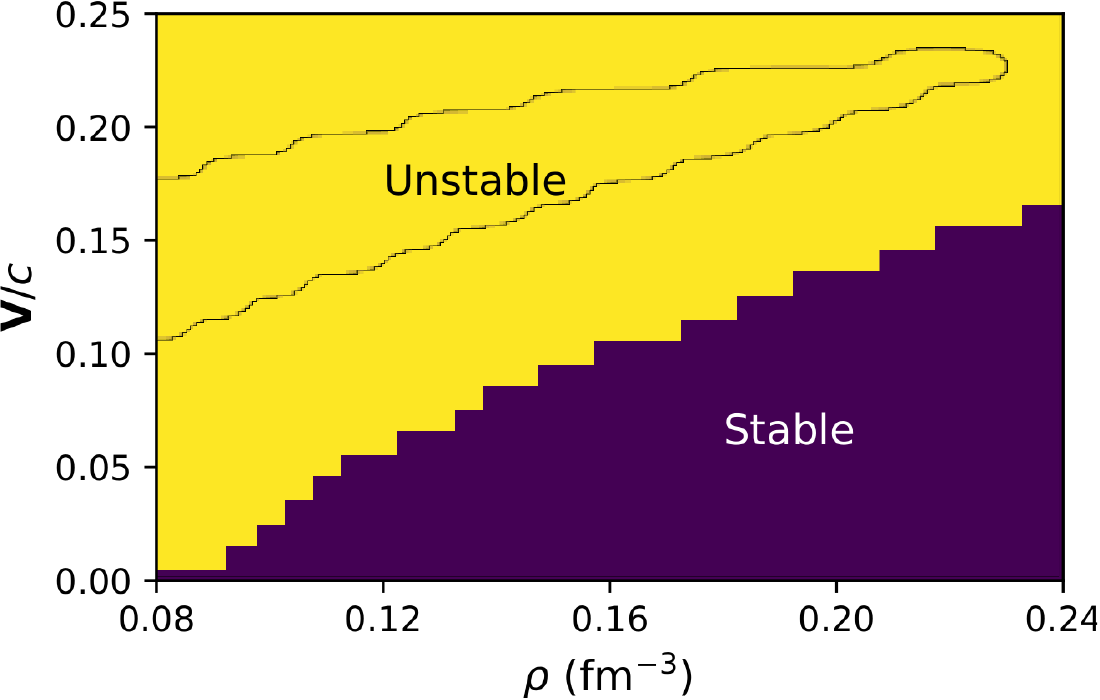}
	\caption{(a) Rotonic structure in the dispersion of linear excitations of a uniform state with relative superfluid velocity $\mathbf{v}_{pn}=2\mathbf{V}=0.15$ c. The roton minimum belongs to the lowest, positive energy branch and takes place at $k_r=0.31$ fm$^{-1}$. Simultaneously there exist a rotonic structure in the highest, negative energy branch that presents a maximum.
		(b) Range of unstable uniform states (yellow color) in the presence of relative superfluid velocity $\mathbf{v}_{pn}=2\mathbf{V}$. The region encircled by the continuous line indicates the presence of unstable modes with wavenumber $k=0$.}
	\label{fig:imap}
\end{figure}
\subsection{Rotonic structures and instabilities}

Roton excitations were introduced by Landau in order to explain the varying dynamics of superfluid Helium with temperature \cite{Landau1941}. While long-wavelength modes (phonons) were the main contribution to the superfluid excitations near zero temperature, short-wavelength modes around a minimum in the dispersion curve (rotons) were the dominant type of excitations at slightly higher temperatures (see, e.g., Ref. \cite{Feynman} for a microscopic characterization of roton excitations in liquid Helium). Differently from phonons, and due to the minimum in the dispersion diagram, the group velocity of rotons vanishes, and the roton modes are localized in a region of the order of the inverse wavenumber $k_r^{-1}$ corresponding to the local energy minimum. On the other hand,
rotons present the lowest phase velocity of the collective superfluid excitations, and so, following the Landau's criterion, they are expected to mark a threshold for the critical velocity of an external probe capable to excite the superfluid. Still, this collective excitation threshold competes with the fermionic critical velocity associated with the pair-breaking excitations \cite{Combescot2006}.

As can be seen in Fig. \ref{fig:imap}(a), which zooms in the bottom inset of Fig. \ref{fig:freq016} to represent the lowest positive-energy branch (solid line) of the dispersion at $\rho=0.16$ fm$^{-3}$, the two-superfluid system of the inner core develops rotonic structures in the presence of relative superfluid velocity.
For $\mathbf{V}=0.075 \,c$, the minimum phase velocity $\omega/k\approx 0.0085\,c$ is reached at $k\approx 0.4$ fm$^{-1}$, nearby the roton minimum at $k_r$, and it is given by the tangent to the dispersion curve passing through the origin (dashed line). The tangent decreases for increasing relative superfluid velocity. Notice that although the negative branches of the dispersion [not shown in Fig. \ref{fig:imap}(a)] are not the mirror reflection of the positive branches with respect to the zero energy axis (due to the effect of combined population imbalance and relative motion), they follow a symmetric trend by decreasing their negative phase velocities. For high enough relative superfluid velocity the two mentioned dispersion branches intersect at their respective rotonic regions, marking the threshold for dynamical instabilities. 
Beyond this threshold, the growth of the unstable (rotonic) modes from small perturbations during the time evolution of the system can eventually produce the decay of the stationary state.

The whole instability region predicted by the present model is depicted in the density--velocity map of Fig. \ref{fig:imap}(b). 
The system shows a relative velocity threshold for the onset of instabilities that increases (approximately) linearly with the total density. Most of the unstable region is dominated by rotonic instabilities, and only the region encircled by the continuous line contains unstable modes at zero quasimomentum. In this way, our results generalize previous work on dynamical instabilities in the two-superfluid outer core \cite{Andersson2004,Kobyakov2017disp}, where only phonon instabilities were identified.

\subsection{Spectrum of linear excitations}

\begin{figure}[t]
	\flushleft{ \bf{(a)}}
	
	\centering
	\includegraphics[width=\linewidth]{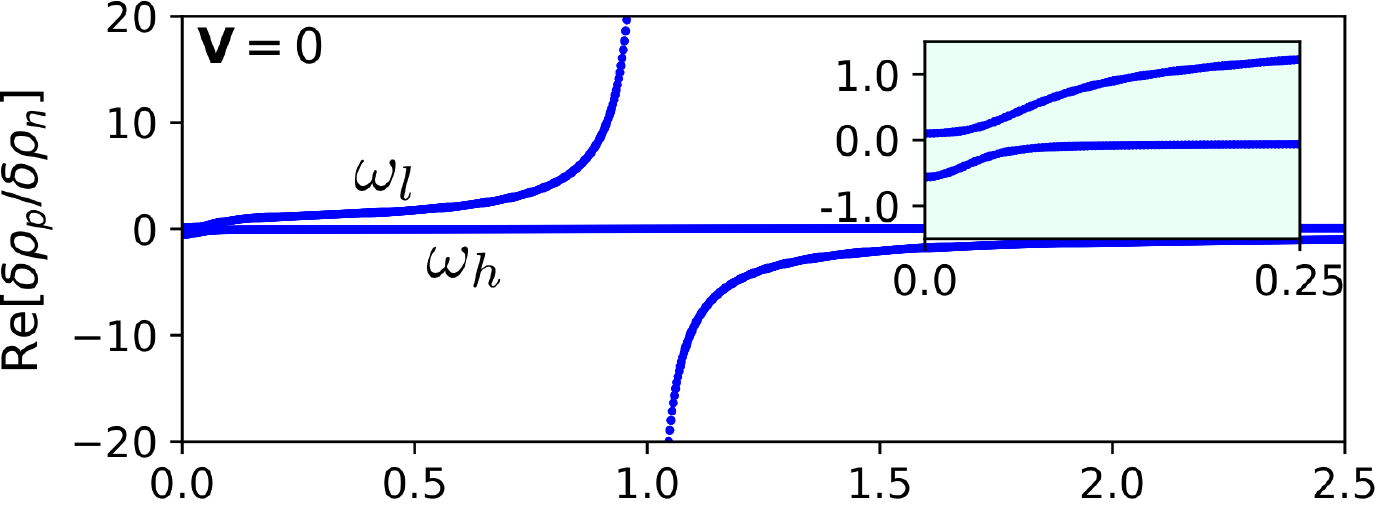}\\
	\vspace{-0.3cm}
	\includegraphics[width=\linewidth]{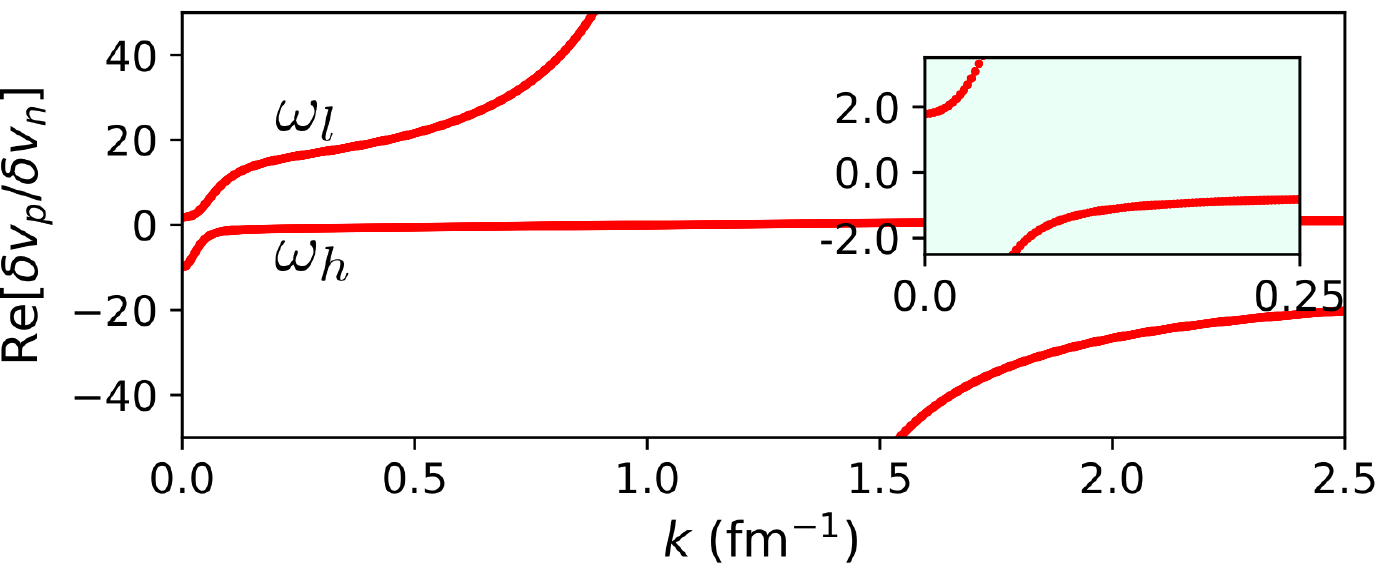}\\
	\flushleft{ \bf{(b)}}
	\includegraphics[width=\linewidth]{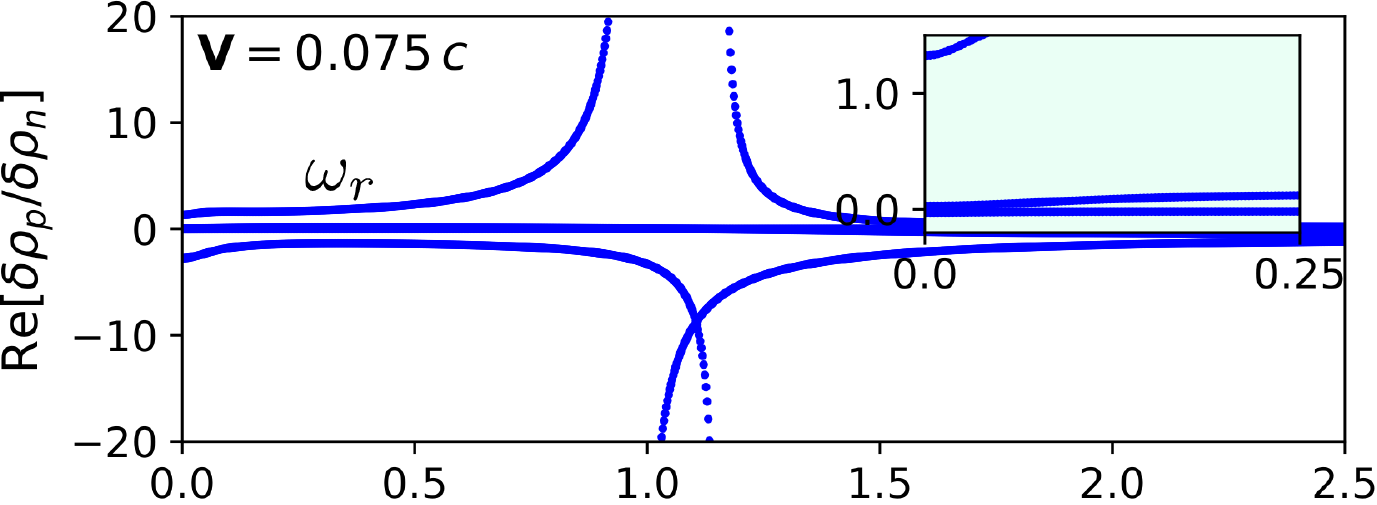}\\
	\vspace{-0.3cm}
	\includegraphics[width=\linewidth]{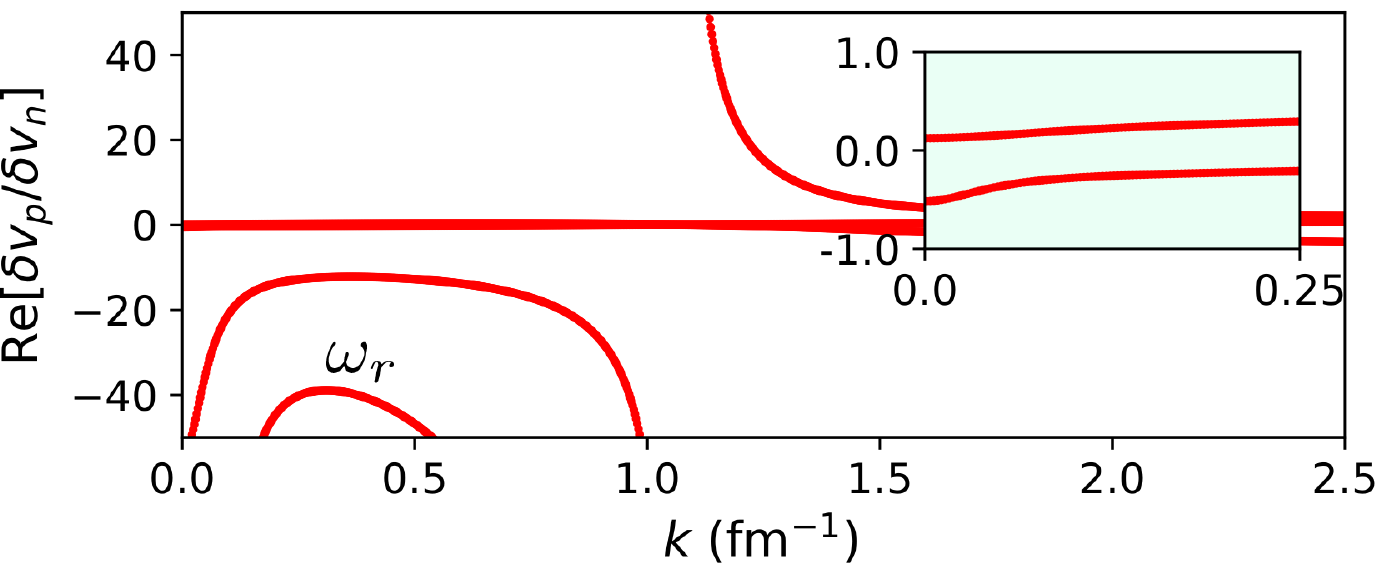}
	\caption{Excitation modes of constant density states with nuclear density $\rho=0.16$ fm$^{-3}$, and relative velocity $\mathbf{V}=0$, 
		top panels (a), and $\mathbf{V}=0.075 \,c$, bottom panels (b), corresponding to the dispersion relations depicted in the top panel of Fig. \ref{fig:freq016}. 
		The label $\omega_l$ (respectively $\omega_h$) indicates the correspondence of the mode branch with the lowest (resp. highest) energy branches of the dispersion; analogously, the label $\omega_r$ refers to the rotonic branch of the dispersion.
		The continuation of this latter branch in the velocity component panels at high $k>1.2$, at high positive values $\delta \mathbf{v_p}/\delta \mathbf{v_n}>0$, is not captured by the graph scale.
		Since the systems are dynamically stable, all modes are chosen to have only real components. The insets zoom in the low wavenumber range in order to see the in- and out-of-phase character between neutron and proton modes.  }
	\label{fig:mod016}
\end{figure}
A common feature of the spectrum that appears at both zero and non-zero relative velocity is the crossing  of in- and out-of- phase modes; see e.g. Fig.  \ref{fig:mod016}, where the crossing takes place in the range of $k\in [1,\,1.2]$ fm$^{-1}$ at $\rho=0.16$  fm$^{-3}$.
As predicted by the simplified phononic models represented by Eqs. (\ref{eq:sound0}) and (\ref{eq:sound1}), in the limit $k\rightarrow 0$ the spectrum is made of proton to neutron in-phase modes at low energy, and out of phase modes at high energy. This fact reflects the 
phase locking character of both density and entrainment couplings, which take negative values in the mentioned limit, $g_{np}(k=0)=(\partial \mu_n/\partial \rho_p)<0$ (although $g_{np}(k=0)>0$ for $\rho>0.21$ fm$^{-3}$) and $\varrho_{np}<0$. 
However the coupling terms in Eq. (\ref{eq:linearB}) are modified for varying wavenumber and nonzero relative velocity. In particular, the density coupling becomes [see Eq. (\ref{eq:Eij})] $g_{np}-\alpha m \mathbf{V}^2/2=(\partial \mu_n/\partial \rho_p) + \vartheta_{np}(\hbar k)^2/2m-\alpha m \mathbf{V}^2/2$, and even at zero velocity $\mathbf{V}=0$ the positive $k$ dependent term competes with the negative chemical potential derivative to give a definite sign to the coupling. 
For high enough wavenumber $k^2>2m|\partial \mu_n/\partial \rho_p|/(\hbar^2 \vartheta_{np})$, the density coupling changes the sign and the spectrum modes reverse their phase character (either in-phase or out of phase). Due to the entrainment coupling, this mode crossing is not simultaneous for velocity and density components, and there exist a transition region for a small range of wavenumbers where the velocity perturbations do not follow the phase character of the density perturbations, that is, one can observe modes having in-phase velocity components and out-of-phase density components. 
However, in general, in the range of wavenumbers where the crossing takes place, the collective excitation energies are much higher than the pair-breaking limit and the corresponding modes are expected to be damped.

More relevantly, the presence of relative velocity breaks the degeneracy between excitation branches of the spectrum, reflecting the underlying asymmetric 
 dispersion. When $\mathbf{V}=0$ (see the top panels of Fig. \ref{fig:mod016}), the spectrum shows only two curves, each of them corresponding to two overlapped branches of linear excitations with opposite phase velocities. The modes labeled by $\omega_l$ (respectively  $\omega_h$) correspond to the lowest (resp. highest) energy branches of the dispersion. This correspondence is reversed after the before mentioned mode crossing.
 When $\mathbf{V}\neq 0$, as can be seen in the bottom panels of Fig. \ref{fig:mod016}, the mode branches split and two new (with respect to the static case) curves can be seen to emerge. Two of the mode branches are close to pure neutronic excitations, since
$\delta \mathbf{v_p}/\delta \mathbf{v_n}\sim \delta \rho_p/\delta \rho_n\sim 0$ (see the insets). 
The key difference introduced by the relative velocity is the fact that one of branches contains modes (labeled by $\omega_r$), belonging to the rotonic region of the dispersion (cf. Fig. \ref{fig:imap}), that combine in-phase 
density perturbations and out-of-phase velocity perturbations, in spite of the underlying density plus entrainment coupling in this wavenumber range, which tends to lock in-phase all the mode components. This feature is relevant for triggering the dynamical instabilities since the associated rotonic modes occupy the lowest energy region of the spectrum, and therefore they are not expected to be damped by pair-breaking mechanisms.

\begin{figure}[t]
	\includegraphics[width=\linewidth]{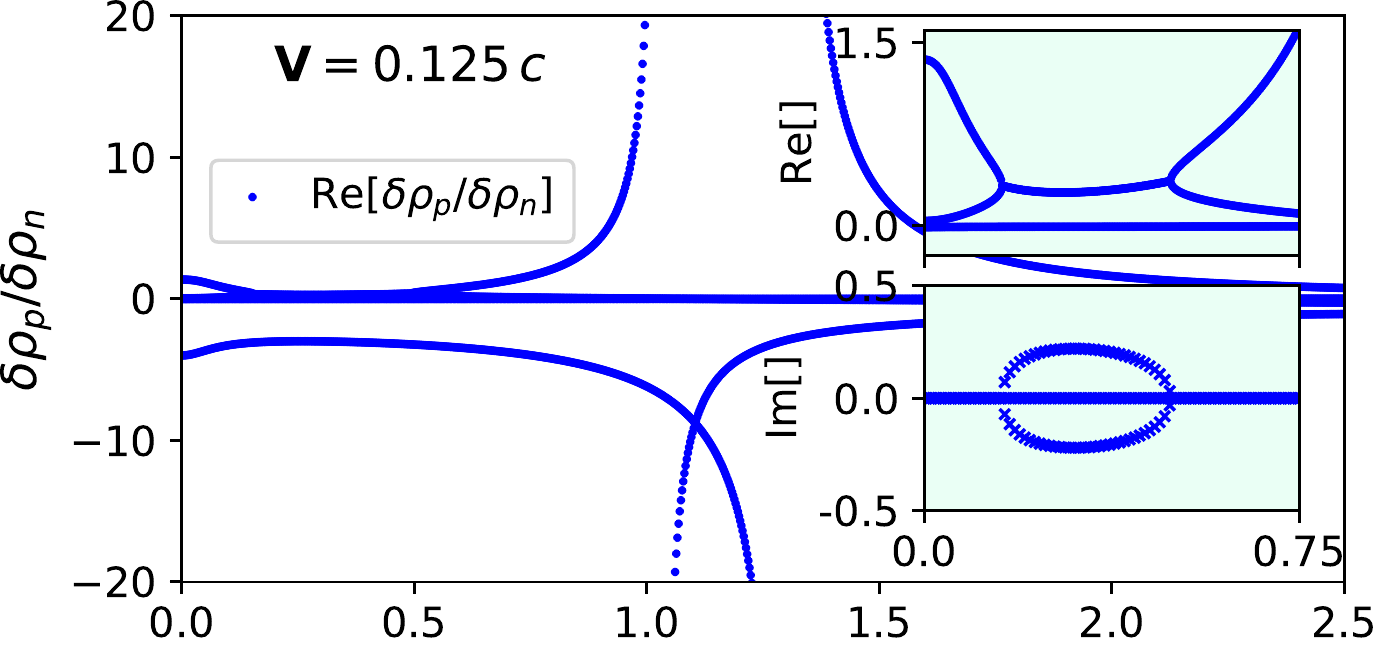}\\
	\vspace{-0.3cm}
	\includegraphics[width=\linewidth]{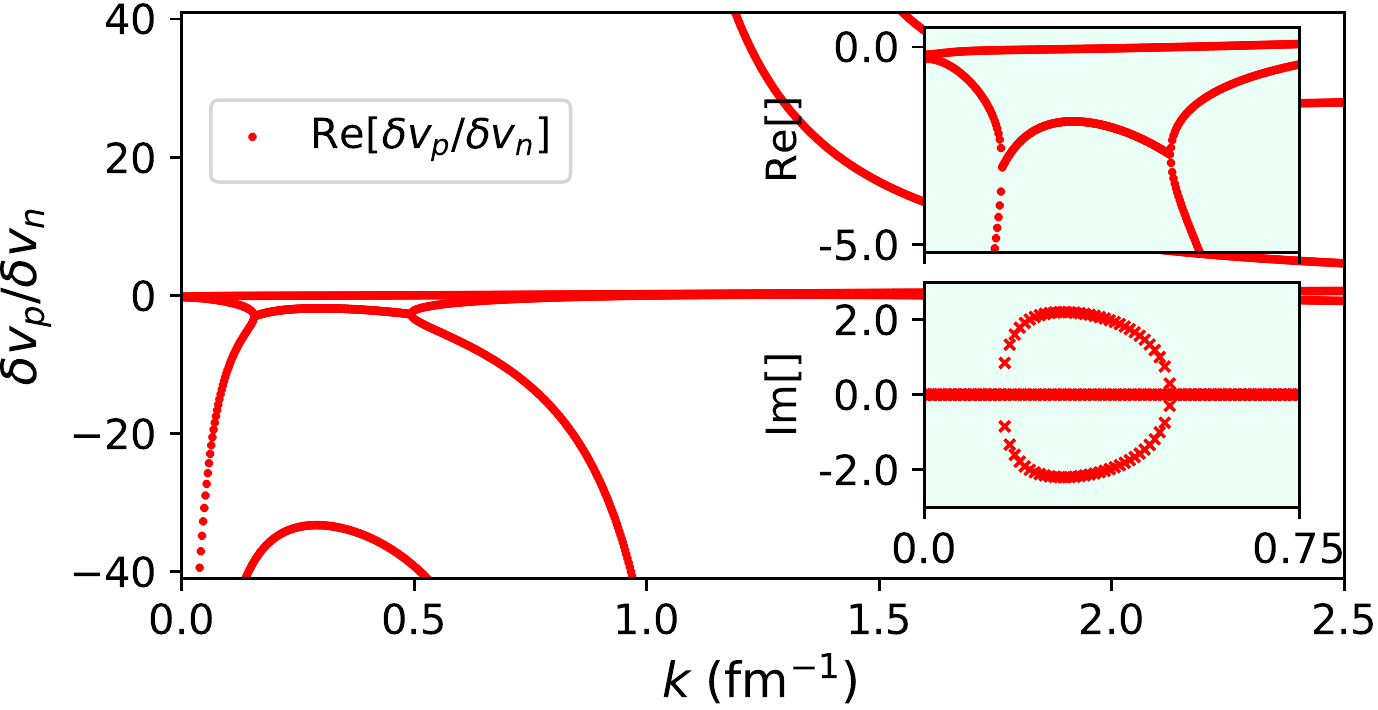}\\
	\caption{Same as Fig. \ref{fig:mod016} for an unstable stationary state at total density $\rho=0.16$ fm$^{-3}$ and relative superfluid motion with $\mathbf{V}=0.125 \,c$. 
		The represented modes correspond to the dispersion relations shown in the middle and bottom panels of Fig. \ref{fig:freq016}. 
		Inside the region of unstable wavenumbers $k\in  [0.16,\,0.49]$ fm$^{-1}$, the modes are described by complex components. The main panels depict the real part of the mode components versus wavenumber, whereas the two insets within each main panel represent their real (top inset) and imaginary (bottom inset) parts.}
	\label{fig:mod016i}
\end{figure}

The onset of instability at high relative velocity can be understood as a resonant process that
occurs at the matching of two excitation frequencies. As can be seen in the bottom panels of Fig. \ref{fig:freq016}, for $\mathbf{V}=0.125 \, c$, the instability region is delimited by the collisions of the two lowest  (in absolute value) energy branches 
of the dispersion. The associated unstable modes, whose Fourier components $U_\mathbf{k}$ are characterized in Fig. \ref{fig:mod016i}, 
show clearly the merging process and the appearance of complex components. 
Notice that with increasing relative velocity, the rotonic mode branch in the velocity panel of Fig. \ref{fig:mod016} (at $\mathbf{V}=0.075\,c$) has shifted its position closer to the horizontal axis to meet the other low energy branch.
These modes involve perturbations that satisfy $\delta \mathbf{v_p}/\delta \mathbf{v_n}\times \delta \rho_p/\delta \rho_n < 0$. Such an opposite relation between the density and the velocity components of the excitation modes, on the top of the dynamical  
equilibrium of coupled superfluids, is unstable (once the relative velocity is beyond a density-dependent threshold), and it is the expected cause of decay of the stationary configuration. 

For each of these modes with wavenumber $\mathbf{k}$, there exist another degenerate unstable mode with opposite wavenumber  $-\mathbf{k}$, such that their linear combinations give rise to standing waves of typical wavelength $\sim 2\pi k_r^{-1}$, since the instability appears in the rotonic region of wavenumbers. The standing wave pattern
involves in-phase density modulations in the proton and neutron superfluids that are opposite (over-density versus under-density) at both sides of each of the standing wave nodes. But simultaneously, the velocity excitations are out of phase for protons and neutrons, so that one superfluid "accelerates" while the other "decelerates" in between the standing wave nodes, which become separated by regions of either increasing or decreasing relative velocities; the growth of these excitations can eventually evolve into quantized superfluid vortices.
 However the linear analysis cannot predict the outcome of the nonlinear time evolution of the system. Therefore,
whether these density modulations can grow and the associated vortices can emerge, eventually giving rise to a new, stable stationary configuration is beyond the 
present analysis, and it will be studied elsewhere.

\section{Conclusions}

In the search for physical phenomena that could be responsible for the observed timing anomalies of pulsars, previous works have explored the two-stream instabilities originated by unstable phonon excitations in the superfluid interior of neutron stars \cite{Andersson2004}. The present work contributes to this exploration in the outer core of neutron stars, and generalizes the previous findings by showing that rotonic excitations may also lie at the origin of dynamical instabilities, and could eventually lead to emergent vorticity along with modulations of the superfluid density. Since these collective modes are among the lowest energy excitations, they can remain undamped by pair-breaking effects.

By means of an effective nuclear interaction Skyrme SLy4, and in the presence of dynamical entrainment (computed with the same Skyrme force), we have shown from the linear analysis of the two-superfluid hydrodynamics that
 the rotonic structures are originated at intermediate wavenumbers ($\sim 0.4$ fm$^{-1}$) in the presence of relative motion between neutron and proton superfluids. Either rotational lag between the two superfluids or star precession could be at the origin of such relative motion, and hence of the associated instabilities \cite{Andersson2004,Glampedakis2008,Glampedakis2009,Andersson2013}. 
  We have found rotonic instabilities above relative superfluid velocities of $\sim 0.1$ c at $\rho$=0.16 fm$^{-3}$, a velocity threshold that increases 
  monotonically with the nuclear density. The associated unstable modes present in-phase superfluid densities and out-of-phase superfluid velocities, and have a fast exponential growth in the linear regime (typical times of above twice the proton plasmon period have been found at saturation density). The final fate of these modes in the nonlinear regime is beyond the scope of this work. The effect of alternative nuclear-interaction forces on roton instabilities presents also an interesting prospect of further study.
  Another point to be investigated is the role of the entrainment in the roton instabilities. As far as the entrainment can be expressed in terms of the isovector effective mass \cite{Allard2019}, it would be instructive to test the impact of energy functionals with different effective masses on the appearance of rotons.

 Our simplified non-relativistic model (see Ref. \cite{Haber2016} for a relativistic approach) does not include the Landau damping associated to the electron gas \cite{Jancovici1962,Kobyakov2017disp}, the
mutual friction between superfluids from underlying vorticity \cite{Alpar1984,Mendell1991},  
the viscosity from a normal fluid \cite{Andersson2019}, nor magnetic-field effects \cite{Eysden2018,Kobyakov2018}. These features will be considered in future works.

\begin{acknowledgments}
	The authors are grateful to Nicolas Chamel and Dmitry Kobyakov for their feedback on an early version of the manuscript.
	X.V. was partially supported by Grant FIS2017-87534-P from MINECO and
	FEDER.
	X.V. and J.A.G.G. also thank the support from the State Agency for Research of the Spanish Ministry of Science and Innovation through the Unit of Excellence Maria de Maeztu 2020-2023 award to the ICCUB (CEX2019-000918-M).
\end{acknowledgments}

\bibliography{neutron_star}
\bibliographystyle{apsrev4-1}

\end{document}